# Integration of Flexible Web Based GUI in I-SOAS

Z. Ahmed[a,b], V. Popov[b,c]

[a] *University of Wuerzburg, Germany*
[b] *Vienna University of Technology, Austria*
[c] *Technical University Sofia, Bulgaria*

**Abstract**

It is necessary to improve the concepts of the present web based graphical user interface for the development of more flexible and intelligent interface to provide ease and increase the level of comfort at user end like most of the desktop based applications. This research is conducted targeting the goal of implementing flexible GUI consisting of a visual component manager with different components by functionality, design and purpose. In this research paper we present a Rich Internet Application (RIA) based graphical user interface for web based product development, and going into the details we present a comparison between existing RIA technologies, adopted methodology in the GUI development and developed prototype.

**General Terms**: RIA, GUI, I-SOAS

**Keywords**: Graphical User Interface, Semantic Web, Rich Internet Application

## 1. Introduction

Currently, the World Wide Web is mainly based on documents written in Hypertext Markup Language (HTML), a markup convention that is used for coding a body of text interspersed with multimedia objects such as images and interactive forms. HTML has many limitations for searching and processing data from it. The main objective of this research is to provide prototype environment for web based intelligent graphical interface.

The Graphical User Interface (GUI) must consist of a visual component manager with different components by functionality, design and purpose. The GUI must be flexible and dynamic, to enable moving of objects, changing colors and fonts, choosing between components in the interface, changing user details, extract and view specific information from the database, communicate with other users in the system. Moreover the GUI can also be saved and restored after user logoff and login.

This research paper is divided into six sections. In section 2 we present RIA frameworks in details for better communication between the graphical user interface and the server. In section 3 we present some available RIA technologies and perform a comparison to choose one the best RIA technology. In section 4 we explain the association of this research work with I-SOAS. We also discuss the implemented design methodology, application flow and structure of GUI of developed prototype version of I-SOAS. In section 5 we briefly present the prototype and in section 6 we conclude the discussion.

## 2. RIA Development

Rich Internet Application (RIA) is web application

that is similar to desktop applications, mostly delivered by web browser plug-ins. Some of them are running independently via sandboxes or virtual machines. Present RIA technologies include Adobe Flex/AIR, JavaFX, Microsoft Silverlight etc. with the new versions of HTML and XHMTL appeared frameworks like jQuery, based on JavaScript and HTML DOM (Document Object Model) interfaces. The biggest developing companies in this sector are committed to enable RIA developers to innovate and create the next generation of Web and desktop applications. Usually an appropriate RIA framework is required to run the application, and needs to be installed using the computer's operating system before launching the application. This software is typically responsible for downloading, updating, verifying and executing the RIA.

To work with a system, users have to be able to control the system and assess the state of the system, also and to get results from it in the proper format. RIA present a new way of developing web based Graphical User Interface (GUI) including new features and more capabilities than the standard web pages. The design and the good usability of the user interface is a significant part of the research process of the project. The Intelligent User Interface (IUI) is intended to provide innovative implementations, systems and technologies and apply the new ideas about how people and computers could "intelligently" interact. Consequently, the design of effective and efficient human-computer interfaces becomes ever more critical to overall system performance. Main issues addressed by intelligent user interface research are the following:

- How can interaction be made clearer and more efficient?
- How can interfaces offer better support for their users' tasks, plans, and goals?
- How can information be presented more effectively?
- How can the design and implementation of good interfaces be made easier?

The design of a user interface affects the amount of effort the user must expend to provide input for the system and to interpret the output of the system, and how much effort it takes to learn how to do this. Usability is the degree to which the design of a particular user interface takes into account the human psychology and physiology of the users, and makes the process of using the system effective, efficient and satisfying.

Usability is mainly a characteristic of the user interface, but is also associated with the functionalities of the product and the process to design it. It describes how well a product can be used for its intended purpose by its target users with efficiency, effectiveness, and satisfaction, also taking into account the requirements from its context of use.

**3. RIA Technologies**

RIA is the next generation of web and desktop applications and combines the best of the user interface of desktop applications plus low-cost deployment of Web applications and rich interactive multimedia communication. It eases the design and development of Intelligent User Interface applications with new ideas and a lot of features of the IDE and for the developed application. In the beginning of the research we explored the present RIA frameworks and RIA development approaches. Because of the fact it cannot be compared with all frameworks, we chose the most wide-spread and stable ones: Flex/AIR (Adobe), Silverlight (Microsoft) and Java FX (Sun). We compared them in several trends: world-wide usage, speed, interface capabilities, IDE and development easiness and platform dependency. Also a valuable requirement is the integration with java because of that the project I-SOAS is being built on the java server platform. We summarized the results in the following table 1.

We chose Flex because of some big benefits of it ahead of the other two frameworks. First of all it is based on flash and it runs in plash player plug-ins, so it is very easy to user without additional plug-ins, because most of the computers have installed plash player in their browsers. Also both it and JavaFX have advantage that they can run on any operating system, beating Silverlight that it runs only on Windows platforms with .Net support. Flex has very good performance and have video and graphic acceleration, JavaFX don't have such. Flex builder have a lot of components for use in the developed application and is better IDE than the other two's. After the research we concluded that it is best for the communication between Flex and Java server, to use BlazeDS server plug-in, which is based of the AMF (Action Message Format). AMF is free and have server side includes for a lot of languages, including Java. It is binary format based loosely on the Simple Object Access Protocol (SOAP). It eases the management of the requests to the server-side application, using Remote Procedure Call (RPC) and the returned result is expressed as an ActionScript Object.

Table 1. Comparison RIA Technologies

| Feature | Flex | Silverlight | Java FX |
|---|---|---|---|
| IDE GUI | yes | yes | no, source code only, new script language |
| Project UI declarations | XML based (MXML) | XML based (MXML) | declarative expressions |
| OS integration/cross-platform | yes (Flash or FlexAIR) | Windows only (requires .Net support) | yes (Java runtime library) |
| Server side integration | object based, AMF | object based, AMF | AJAX, java services |
| Worldwide usage | best | poor | good |
| Loading time / Boot | fast | good | slow |
| 3D | good (HW support) | good (HW support) | poor (no HW support) |
| Components & Tools | better | good | good |
| Component integration with OS | good | bad (no cross platform) | good |

## 4. I-SOAS

The project Intelligent Semantic Oriented Agent based Search (I-SOAS); is a project proposed for the development of a web based flexible graphical user interface and intelligent search in Product Data Management Systems [1, 2]. The conceptual architecture of I-SOAS is categorized into four different sequential iterative parts i.e. Intelligent User Interface (IUI), Information Processing (IP), Data Management (DM) and Data Representation (DR), as shown in Figure 1. In this research paper, we are not going into the detail of any of the four parts of I-SOAS but IUI, the main graphical user interface [3] as the development of IUI using RIA technologies is the main goal of this research (presented in this research paper).

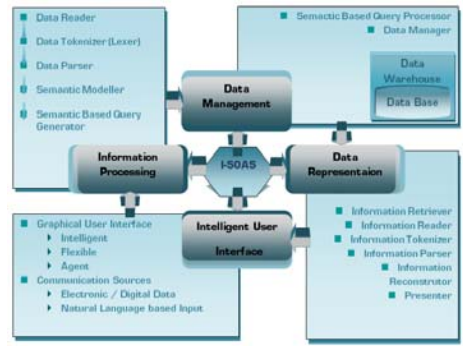

Figure.1. I-SOAS Conceptual Architecture

### 4.1. Design Methodology

The project structure consists of 3 basic parts: Flex files (SWF flash files), Java classes and a database. In future it is also planned to support external database, so the user may work in different areas from one place. Every part has different ways of working and developing process. These parts are integrated as one application like product line applications.

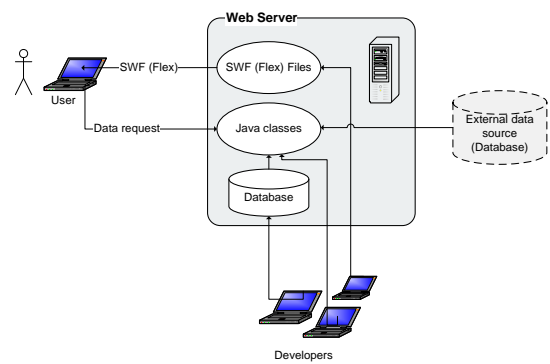

Figure.2. I-SOAS Project Design Methodology

The working process includes a request from the user to open the page for which the client browser receives the SWF files (flex compiled application) from the web server. Every data request is managed by the BlazeDS library (java remote include classes framework). After that the data is retrieved from the database and is sent as SWF data (compiled flex / ActionScript objects) to the interface. The communication with the database is made only with the java classes, as shown in Figure 2.

### 4.2. Layer Communication

The communication between layers is present in the following Figure 3. The communication is managed by AMF (Action Message Format). Its data format is binary based using the Simple Object Access Protocol (SOAP). It is versioned with server side includes for most of the programming languages.

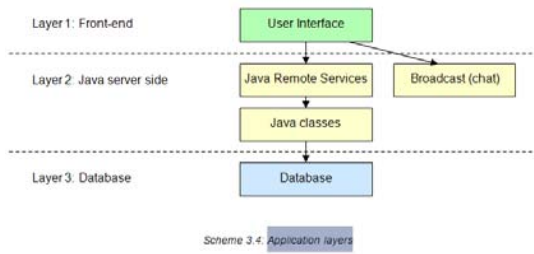

Figure.3. Application layers process

AMF eases the management of the requests to the server-side application, using Remote Procedure Call (RPC). With this technology, methods and classes at the server side (Java) can be called and be sent easily direct from the front-end application (Flex), without any additional code. There is only a configuration file mapping the class names. The result from the execution is expressed as an ActionScript Object. For now best Java based Server Side Includes (SSI) is BlazeDS developed by Adobe. The request is asynchronous, so the application does not wait for the reply from the server and result can be displayed at any time. When a data request is to be executed, e remote call is made from the user interface in the remote services' (via the server side includes) class members and the result is sent as an object of a Java class. If there have to extract data from the database (for ex. Search, login / logout …) a permanent connection to the database is created and the query and data is processes via the Java-MySQL connector add-on. The retrieved data is processed in the remote services and returned to the user interface module as object of a specified class type. When a user is logged on, he/she is subscribed for the BlazeSD broadcasting service and can send and retrieve messages in the chat component.The remote services is used in the following cases: log-in or register user, loading and saving components' saved states, searching phrase or SQL query, loading and saving user interface settings, get actions log and chat. Some of the components do not use the remote services, because they do not require user data to work and can be accessed without login.

*4.3. Application Flow*

The main flow describes almost all flow of the available features in the system. It describes the begin point, how to proceed to search in I-SOAS and the work with the components. The work flow is presented in Figure 4.

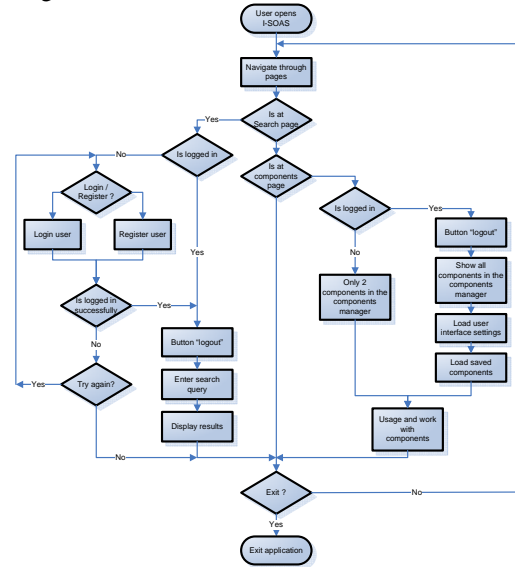

Figure.4. Application flow

The interface is made like that so the user can feel free to navigate from anywhere he/she wants to anywhere else. The data and the state of the pages are preserved and shown again after going back to the page. If the user wants to search for something, he/she must login first and after that the search is available. The "Logout" button is visible only after successful login or register. When the user opens the components page, only 2 components are shown in the component manager if the user is not logged. If is logged all components are shown and after that the user interface settings are loaded and applied. Next if there is saved components, they are restored.

*4.4. GUI Structure*

Front-end is completely developed with Flex. The flex application has main module for page container and by one module for every page in the project. Every component in the application is also separated as shown in Figure 5. There is a menu bar on the top of the page container, which carry out the navigation functionality. If the user is logged-in, a "Logout" button is shown in the right of the menu bar. In the scheme below is displayed the structure of the interface of I-SOAS prototype.

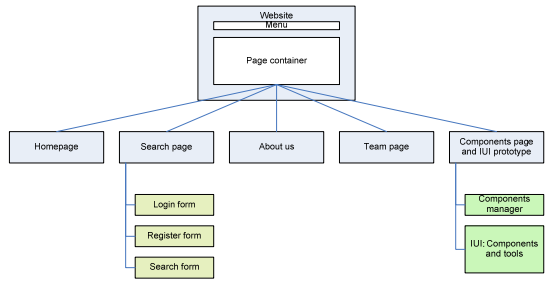

Figure.5. Application GUI structure

Every page opens in the page container and every component in the page can be moved and shown in this container. The page module is created on first use, instead of creating in load time. This speeds up a bit the loading for slow client computers and if user does not open some pages, they will not be created. Once created, the page is not destroyed and created again, but only hidden and shown again. That is how the view and components inside the page are preserved at runtime, as shown in Figure 6.

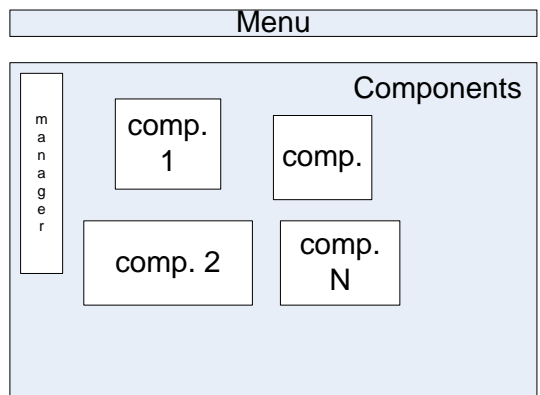

Figure.6. GUI components page structure

Homepage is the main page, Search page provides search and login options, About us page – contains contacts, Team page – contains a list of out team worked on this project and Components page – contains the prototype of the IUI with a component manager and user area. If the user is not logged in, only the components that do not require user data to work are available. If the user is logged-in, all components are available in the component manager. Also if there is saved component's state from the previous session, the saved state is restored and the user interface settings are loaded and applied. There is a counter of the displayed components in the page in the bottom of the components manager window. Components are organized in own windows. Every component window, including the component manager's window, can be moved in the user area. The component windows have animations for showing and hiding events, for showing status message in the bottom of the window and animations for the tool tips. The components page's scheme is illustrated in figure 6. All components are visual classes based on one main class, called MovablePanel, which is based on the Panel component in Flex framework. This class is responsible for the ability of the component to interact as a desktop window. It enables the movement with mouse dragging, adds close button and handles the connection between the component and the user interface. Only the clock component is based on the Label visual component in Flex framework.

## 4. Involved Technologies

We used following software technologies: Flex Builder 3 IDE, Eclipse IDE 3.4, Java SDK, Apache Tomcat 6 server, BlazeDS 3.2 – java remoting services plug-in, MySQL 5 Database Server and MySQL Connector 3.0. for the development of I-SOAS prototype.

## 5. Prototype

This is IUI of I-SOAS Web Application as shown in figure 7, capable of doing following tasks

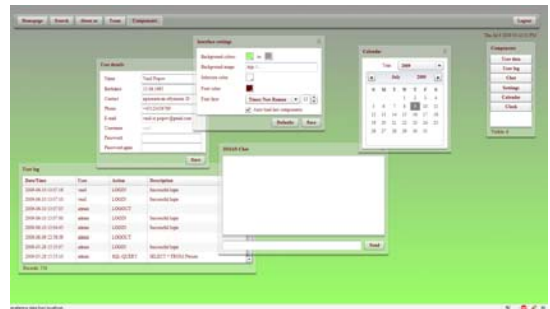

Figure.7. I-SOAS GUI Prototype

1. Providing a complete user authentication mechanism which let new user register and old user login by providing valid user name and password.
2. Providing a Visual components manager for component management.
3. Providing options to user to redesign a new graphical user interface by adding deleting and arrange components

4. Providing option to save newly build graphical user interface.
5. Providing Visuals effects in graphical user interface development.
6. Providing options to search records by entering natural language or SQL queries

## 6. Conclusion

In this research paper we have discussed an approach to integrate intelligent user interface in a semantic web based application. This interface obeys the assigned tasks and criteria and also we even added additional features and effects to it. It is flexible enough to manage different type of components, different user preferences for design, enable the user to search in two different ways, change between different states and interface settings whether if the user is logged-in or not. The interface has many visual effects and flexible and has very easy changeable design. I-SOAS is Flex and Java based application and uses a relational database with MySQL database server, where stores and manages its data. The front-end (the user interface) communicates with the server and gathers data from the database thorough BlazeDS java plug-in, which is based on the AMF (Action Message Format).

In the beginning of the paper we made a research of the present RIA technologies and compared them to show the benefits and weaknesses of every particular technology framework. Also we researched some of the available PDM applications and compared them to the concept of I-SOAS. We got some ideas from those applications, so to improve the prototype of I-SOAS Intelligent User Interface.

In future we are hoping to focus on depth of architectural details, mathematical or algorithmic experiments and currently available tools and technologies for the development of the project and to improve some modules in it. As we are promoting I-SOAS as a dynamic concept of Intelligent User Interface we will also try to prove the importance of the IUI web integration in different fields in the business process and work, also and in the entertainment area.

## Acknowledgements

We are thankful Vienna University of Technology and Technical University Sofia for giving us the opportunity to work on this project (I-SOAS). We are also thankful to Erasmus for funding Vasil Popov for working this project. We are also thankful to Prof. Dr. Detlef Gerhard for his supervision during the research and development of this project.

## Supplementary Web Links

1. Rich Internet, Last reviewed 06 November 2009, <http://en.wikipedia.org/wiki/Rich_Internet_applicationY
Java, Last reviewed 06 November 2009, <http://www.java.com/en/>
2. Flex, Last reviewed 06 November 2009, <http://www.adobe.com/products/flex>
Silverlight, Last reviewed 06 November 2009, <http://en.wikipedia.org/wiki/Silverlight>, 2008
3. Apache Tomcat, Last reviewed 06 November 2009, <http://tomcat.apache.org/>
4. BlazeDS 3.2, Last reviewed 06 November 2009,<http://opensource.adobe.com/wiki/display/blazeds/Release+Notes>
5. MySQL, Last reviewed 06 November 2009, <http://www.mysql.com/>
6. Action Message Format, Last reviewed 06 November 2009, <http://en.wikipedia.org/wiki/Action_Message_Format>
7. SOAP, Last reviewed 06 November 2009, <http://en.wikipedia.org/wiki/SOAP>